\documentstyle[aps,prl,twocolumn]{revtex}

\def \eps  {\epsilon}
\def \be   {\begin{equation}}
\def \ee   {\end{equation}}
\def \bea  {\begin{eqnarray}}
\def \eea  {\end{eqnarray}}
\def \d    {{\rm d}}
\input{psfig}

\begin{document}
\bibliographystyle{unsrt}

\twocolumn[\hsize\textwidth\columnwidth\hsize\csname@twocolumnfalse%
\endcsname

\title{Semiclassical Trace Formulas for Two Identical Particles} 

\author{Jamal Sakhr\cite{present} and Niall D. Whelan}
\address{Department of Physics and Astronomy, McMaster University,
Hamilton, Ontario, Canada L8S~4M1}

\date{\today}

\maketitle

\begin{abstract}
Semiclassical periodic orbit theory is used in many branches of
physics. However, most applications of the theory have been to systems
which involve only single particle dynamics. In this work, we develop
a semiclassical formalism to describe the density of states for two
noninteracting particles. This includes accounting properly for the
particle exchange symmetry. As concrete examples, we study two
identical particles in a disk and in a cardioid. In each case, we
demonstrate that the semiclassical formalism correctly reproduces the
quantum densities.
\end{abstract}


]

\section{Introduction}\label{introduc}

\noindent Semiclassical physics has experienced a resurgence of
interest, largely due to the work of Gutzwiller \cite{Gutz}, Balian
and Bloch \cite{Balian} and Berry and Tabor \cite{Tabor}. (For
recent reviews see \cite{ChaosBook,BB}.) These works showed that if we
separate the density of states into smooth and oscillatory components,
then the oscillatory part is related to the dynamics of the underlying
classical system via periodic orbits. This complements the earlier
work of Weyl, Wigner, Kirkwood and others who showed that the smooth
component is related to the geometry of the classical phase space.
Actually, the two components are related in a subtle way
\cite{Cartier,howls} since the complete geometry imparts the
full dynamics and vice-versa.

Most of the theoretical work has concentrated on the single particle
density of states, however, there are some notable exceptions,
namely \cite{sommer,weiden,papen1}. In \cite{sommer} the focus is on
the average level density and its extension to systems of identical
particles. Specifically, the authors consider a system of $N$ fermions
in one dimension. Their Weyl formula for fermions works well for
attractive two body interactions, but overestimates the quantum
staircase function when there are repulsive two body interactions. The
author of \cite{weiden} develops a generalization of the canonical
periodic orbit sum for the special case of $N$ interacting spinless
fermions in one dimension. It is assumed the periodic orbits are
isolated and therefore it is most applicable to fully chaotic systems.
The author also considers a system of noninteracting fermions and
writes the many body level density as a convolution integral involving
one body level densities. Finally, we mention \cite{papen1} which
presents an expansion of the periodic orbit sum in terms of the
particle number using ideas from \cite{sommer,weiden}.

Similarly, most of the applications of semiclassical theory have been
to systems which involve only single particle dynamics. Here, we
mention some exceptions.  The authors of \cite{papen2} extend the
study of scars \cite{scarpapers} to classically chaotic few body
systems of identical particles.  A study of the eigenfunctions of an
interacting two particle system can be found in \cite{benet}. The
semiclassical approach to the helium atom, which can be understood as
two interacting electrons in the presence of a helium nucleus, has
been studied in \cite{helium}. We also mention the novel applications
of semiclassical theory to mesoscopic physics \cite{Richter}. For
example, orbital magnetism has been studied semiclassically for
diffusive systems in \cite{ullmo1} and for ballistic systems in
\cite{ullmo2,kaori}. For reviews see \cite{houches}.

Ultimately, one would like to study an arbitrary number of interacting
particles in any kind of potential. In the present work, we begin by
exploring the structure of the trace formula for two noninteracting
particles including an examination of the decomposition into
bosonic and fermionic spaces. This sets the stage for the
interacting $N-$body problem to be explored in a subsequent publication
\cite{us}. The method employed here uses the fact that the two
particle density of states is the autoconvolution of the single
particle density of states. Subsequently, we decompose the semiclassical two
particle density of states into three distinct contributions, each of
which corresponds to specific dynamical properties of the system.  Of
particular interest is the contribution which corresponds to two
particle dynamics.

Billiards have served as prominent model systems in quantum
chaos. They combine conceptual simplicity (the model of a free
particle in a box) while allowing the full range of classical
dynamics, from integrable to chaotic. Therefore, as initial
applications of the formalism, we study two noninteracting identical
particles in a disk and in a cardioid. The former problem is
integrable while the second is chaotic so these two examples provide a
direct test of the formalism in the two limiting cases of classical
motion. In both cases, we find the semiclassical formalism does a good
job of reproducing the quantum density of states.

\section{Background Theory}

\subsection{Single Particle Semiclassical Theory}

\noindent In this section, we review the formalism for the
semiclassical decomposition of the single particle density of
states. Let $\left\{\eps_i\right\}$ be the single-particle energies so
that the single particle density of states is
\be \label{singpartdens}
\rho_1(\eps) = \sum_i\delta(\eps-\eps_i),
\ee
where the subscript $1$ indicates that it is a single particle
density. A fundamental property of the quantum density of states is
that it can be exactly decomposed into an average smooth part
and an oscillatory part \cite{Balian}
\be \label{decomp}
\rho_1(\eps) = \bar{\rho}_1(\eps) + \tilde{\rho}_1(\eps).
\ee
There are various approaches for calculating these quantities
\cite{BB}. For example, in systems with analytic potentials, the
smooth part may be obtained from an extended Thomas-Fermi calculation
which is an asymptotic expansion in powers of $\hbar$. In billiard
systems, where the particle is confined to a spatial domain by the
presence of infinitely steep potential walls, the smooth part may be
obtained from the Weyl expansion. In two dimensional billiards with
piecewise smooth boundaries and Dirichlet boundary conditions, the
first three terms of the Weyl expansion is \cite{SW}
\be \label{pave1pbill}
\bar{\rho}_1(\eps) = \left({\alpha {\cal A} \over 4\pi}
- {\alpha^{1/2} \over 8\pi} {{\cal L}\over \sqrt{\eps}}
\right) \theta(\eps) + {\cal K}\delta(\eps) + \cdots 
\ee
where $\alpha=2m/\hbar^2$, ${\cal A}$ is the area, ${\cal L}$ is the
perimeter and
\be \label{curvterm}
{\cal K} = {1 \over 12\pi} \oint\d l \kappa(l) + {1 \over 24\pi} 
\sum_i {{\pi^2 - \theta_i^2} \over \theta_i}
\ee
is the average curvature integrated along the boundary with
corrections due to corners with angles $\theta_i$. The oscillating
part is obtained from semiclassical periodic orbit theory, and in
particular the various trace formulas for $\tilde{\rho}_1(\eps)$
of the form \cite{BB}
\be \label{trform}
\tilde{\rho}_1(\eps) \approx  {1 \over \pi\hbar}
\sum_{\Gamma} A_{\Gamma}(\eps) \cos \left ( {1 \over
\hbar} S_\Gamma(\eps) - \sigma_\Gamma {\pi \over 2}
\right ).
\ee
$\Gamma$ denotes topologically distinct periodic orbits,
$S_{\Gamma}(\eps)$ is the classical action integral along the orbit
$\Gamma$. The amplitude ${A}_{\Gamma}(\eps)$ depends on energy, the
period of the corresponding primitive orbit, the stability of the
orbit, and whether it is isolated or non-isolated. The index
$\sigma_{\Gamma}$ depends on the topological properties of each
orbit. For isolated orbits, it is just the Maslov index. For
nonisolated orbits, there may be additional phase factors in the form
of odd multiples of $\pi/4$ which we account for, in a slight abuse of
notation, by allowing $\sigma_\Gamma$ to be half-integer. In the case
of non-isolated orbits, $\Gamma$ denotes distinct families of
degenerate orbits. The amplitude of an isolated orbit
is given by the Gutzwiller trace formula \cite{Gutz}
\be \label{Gutzamp}
     A_\Gamma(\eps) = {T_\gamma (\eps) \over
\sqrt{\left| \det (\tilde{M}_\Gamma - I) \right|}}
\ee
where $T_{\gamma}(\eps)$ is the period of the primitive orbit
$\gamma$, corresponding to $\Gamma$ ({\it i.e.} $\Gamma$ is an integer
repetition of $\gamma$) and $\tilde{M}_{\Gamma}$ is the stability
matrix of that orbit.

\subsection{Quantum Two Particle Density of States} \label{qtpdos}

\noindent Now suppose we have a system of two identical noninteracting
particles. The total Hamiltonian is the sum of the single particle
Hamiltonians and it follows that the energies of the composite system
are just the sums of the single particle energies. The analogue of
(\ref{singpartdens}) is then
\be \label{den1+2q}
\rho_2(E) = \sum_{i,j} \delta (E - ( \eps_i + \eps_j)).
\ee
A useful relation is that the two particle density of states is
the autoconvolution of the single particle density of states:
\be \label{Lpconvthrm}    
\rho_2(E) = \int_{0}^{E}\d\eps\rho_1(\eps)\rho_1(E-\eps)\nonumber\\
          = \rho_1*\rho_1(E),
\ee
as can be verified by direct substitution. In fact, this works even if
the particles are not identical, where the full density is still the
convolution of the two distinct single-particle densities. This would
also apply to a single particle in a separable potential, which is
mathematically equivalent. Rather than encumber the notation to
explicitly allow for this possibility, we defer this discussion to
Appendix A, where some formulas for nonidentical, noninteracting
particles are presented.

We can decompose the two particle density of states for a system of
two identical particles into a symmetric and an antisymmetric density,
\be \label{p=pS+pA}
    \rho_2(E) = \rho_S(E) +\rho_A(E).
\ee
We shall use the terms symmetric/antisymmetric and bosonic/fermionic
interchangeably.  Each partial density may be obtained using a
projection operator onto the relevant subspaces resulting in
\be \label{psaexpandfinE}
    \rho_{S/A}(E) = {1 \over 2} \left( \rho_2(E) \pm {1 \over 2}
\rho_1 \left({E \over 2} \right) \right). 
\ee
We seek semiclassical approximations to these quantum expressions, a
topic which is pursued in the following sections.

\section{Semiclassical Calculations for the Two Particle
System}\label{2ppformalism}

\noindent Decomposing the single particle density into its smooth and
oscillatory components as in (\ref{decomp}) gives a decomposition of
the two particle density of states into three distinct contributions,
\be \label{2pdecomp}
\rho^{\rm sc}_2(E) = \bar{\rho}_1*\bar{\rho}_1(E) +
2\bar{\rho}_1*\tilde{\rho}_1(E) + \tilde{\rho}_1*\tilde{\rho}_1(E).
\ee
The first term is a smooth function of energy since the convolution of
two smooth functions results in a smooth function. This is followed by
a cross term and finally by a purely oscillating term. The cross term
is also an oscillating function. At first, this may seem incorrect
since the convolution of a smooth function with an oscillating
function usually yields a smooth function. As we will show, the
oscillatory nature of the cross term is due to contributions from the
end-points of the convolution integral. Physically, the smooth term
does not depend on dynamics since it corresponds to the Weyl formula
in the full two-particle space. The cross term depends only on single
particle dynamics because it corresponds to the situation where one
particle is stationary and the other particle is evolving dynamically
on a periodic orbit. It is only the last term which contains two
particle dynamics in the sense that both particles are evolving
dynamically on periodic orbits. Hence, we will refer to the last term
as the dynamical term.

We find a general expression for $\tilde{\rho}_1*\tilde{\rho}_1(E)$ by
substituting a generalized trace formula for $\tilde{\rho}_1(E)$ and
then evaluating the resulting convolution integral using the method of
stationary phase. Using (\ref{trform}), the dynamical term can be
written as
\bea\label{2pflucparts}
&&\tilde{\rho}_1*\tilde{\rho}_1(E) \approx {1 \over {(\pi
\hbar})^2} \sum_{\Gamma_1,\Gamma_2}\int_{0}^{E}\d\eps\;
A_{\Gamma_1}(\eps) A_{\Gamma_2}(E - \eps)
\nonumber \\
&& \cos \left ( {1 \over \hbar} S_{\Gamma_1}(\eps) -
\sigma_{\Gamma_1} {{\pi} \over 2} \right )  \cos \left ( {1 \over
\hbar} S_{\Gamma_2}(E - \eps) - \sigma_{\Gamma_2} {{\pi} \over
2} \right). 
\eea
To evaluate this asymptotically, we should include all critical points
in the integration domain. Specifically, this integral has a
stationary phase point within the integration domain and finite valued
endpoints.  We shall show that the stationary phase point corresponds
to the situation where both particles are evolving dynamically with
the energy partitioned between the two particles in a prescribed
way. The endpoint contributions must be evaluated at energies such
that one of the particles has all of the energy while the other has no
energy. However, this contradicts our assumption that both particles
are evolving --- this is the definition of the dynamical
term. Moreover, if we were to evaluate this contribution, the result
would be meaningless since it involves using the trace formula at zero
energy where it is known to fail. So we shall omit the contributions
from the endpoints; this is discussed more fully in section
\ref{spurious_explained} and in Appendix B, as well as in reference
\cite{spurious}.

Hence, we evaluate the integral in (\ref{2pflucparts}) using only the
stationary phase point. To leading order, we can extend the
integration limits over an infinite domain. Writing the cosine
functions as complex exponentials yields four integrals; the first is
\bea\label{I1sp}
&&\int_{-\infty}^{\infty}\d\eps A_{\Gamma_1}(\eps)
A_{\Gamma_2}(E - \eps) \exp \left({i\over\hbar}
\left(S_{\Gamma_1}(\eps) + S_{\Gamma_2}(E - \eps)\right)\right)
\nonumber \\ 
& \approx & A_{\Gamma_1}(E_0) A_{\Gamma_2}(E - E_0)
\sqrt{2\pi\hbar \over \left|\Upsilon (\Gamma_1, \Gamma_2, E)
\right|}\nonumber \\
&& \exp\left({i\over\hbar} (S_{\Gamma_1}(E_0)+S_{\Gamma_2}(E - E_0))
+i\nu{\pi\over 4}\right)
\eea 
where 
\bea\label{secderS}
\Upsilon (\Gamma_1,\Gamma_2,E) & = & \left. \left( {\partial^2
S_{\Gamma_1}(\eps) \over \partial \eps^2}  + {\partial^2
S_{\Gamma_2}(E - \eps) \over \partial \eps^2} \right)
\right|_{E_0} \nonumber\\
\nu & = & {\rm sign} \left(\Upsilon(\Gamma_1,\Gamma_2,E)\right).
\eea
$E_0$ is determined from the stationary phase condition 
\bea\label{eqperiods}
\left. \left({\partial S_{\Gamma_1}(\eps) \over \partial
\eps}  + {\partial S_{\Gamma_2}(E - \eps) \over \partial
\eps} \right) \right|_{E_0}&=&0 \nonumber\\
\Longrightarrow T_{\Gamma_{1}}(E_0)&=&T_{\Gamma_{2}}(E - E_0)
\eea
where we have used the fact that the derivative of the action with
respect to energy is the period. $E_0$ is the energy of particle 1, $E
- E_0$ is the energy of particle 2 and $E$ is the total energy of the
composite system. The saddle energy $E_0$ has a precise physical
interpretation; Eq.(\ref{eqperiods}) says that the energies of the two
particles are partitioned so that the periods of both periodic orbits
are the same. In other words, at $E_0$, we have orbits which are
periodic in the full two particle phase space since after the period
$T$ {\it both} particles return to their initial conditions.

The next integral has the same stationary phase condition as the first
integral and is its complex conjugate. The third integral is
\be\label{I2sp}
\int_{-\infty}^{\infty}\d\eps A_{\Gamma_1}(\eps)
A_{\Gamma_2}(E - \eps) \exp \left(-{i\over\hbar}
\left(S_{\Gamma_1}(\eps) - S_{\Gamma_2}(E - \eps)\right)\right)
\ee
and has no stationary phase point since setting the first derivative
of the action to zero yields the stationary phase condition
\be\label{I2nosp}
T_{\Gamma_{1}}(E_0)=-T_{\Gamma_{2}}(E-E_0).
\ee
The trace formula only involves orbits with positive period, so we
ignore this possibility. The last integral is the complex conjugate of
the third and will also be ignored. 

Adding the contributions from the first two integrals, we arrive at
the two particle trace formula:
\bea\label{oscosc}
&&\tilde{\rho}_1*\tilde{\rho}_1(E) \approx {2\over(2\pi\hbar)^{3/2}}
\sum_{\Gamma_1,\Gamma_2}
{A_{\Gamma_1}(E_0)A_{\Gamma_2}(E - E_0) \over
\sqrt{\left|\Upsilon (\Gamma_1, \Gamma_2, E) \right|}} \nonumber \\
&&\cos\left({1\over\hbar}(S_{\Gamma_1}(E_0) + S_{\Gamma_2}(E- E_0))
-(\sigma_{\Gamma_1} + \sigma_{\Gamma_2}) {\pi \over 2}
+ \nu {\pi\over 4} \right).
\eea
This result possesses the intuitive properties that, other than
factors arising from the stationary phase analysis, the actions and
Maslov indices are additive and the amplitudes are multiplicative.  We
note that this saddle-point analysis fails for the simplest problem in
physics, the harmonic oscillator, where $\Upsilon=0$. This is because
the two-particle harmonic oscillator has a higher degree of symmetry
than we are accounting for here. This is a nongeneric property
specific to the harmonic oscillator. We also stress that we have made
no assumption about the stability or structure of the orbits. They can
be isolated, stable or unstable or come in families. There are also
problems with coexisting isolated orbits and families, such as those
of the equilateral triangle billiard \cite{equilateral,BB}.

Note that the overall $\hbar$ dependence is not multiplicative but
picks up an additional factor of $\hbar^{1/2}$ from the stationary
phase integral. For isolated orbits, the amplitudes $A$ are
independent of $\hbar$ and the expression (\ref{oscosc}) has a
$1/\hbar^{3/2}$ prefactor as opposed to the $1/\hbar$ in the amplitude
of the single particle trace formula. The fact that the $\hbar$
dependence is different implies that the periodic orbits of the full
system come in continuous degenerate families rather than isolated
trajectories, which in turn implies that there exists a continuous
symmetry in the problem \cite{scc}. This is an important point which
we will address in a companion paper \cite{us}. (It was also noted in
\cite{ullmo2}.) Nonetheless, it may be helpful to give a brief
explanation here. Imagine the full phase space periodic orbit $\Gamma$
consists of particle $1$ on a periodic orbit $\Gamma_1$ with energy
$E_0$ and particle $2$ on a distinct periodic orbit $\Gamma_2$ with
energy $E-E_0$. We can define $t=0$ to be when particle $2$ is at some
prescribed point on $\Gamma_2$. Keeping particle $2$ fixed, we can
change the position of particle $1$ on $\Gamma_1$ to generate the
initial condition of a distinct but congruent periodic orbit in the
full phase space. Continuous time translation of the initial condition
on $\Gamma_1$ generates a continuous family of congruent periodic
orbits in the full phase space. Since the time translational symmetry
can be characterized by a single independent symmetry parameter, the
$\hbar$ dependence is ${\cal O}\left(1/\sqrt{\hbar}\right)$ stronger
than for a system with isolated periodic orbits \cite{scc,BB}.

\section{Two Particle Quantum Billiards}\label{2ppbillgen}

\noindent As an application of the formalism developed in section
\ref{2ppformalism}, we consider the quantum billiard problem.
Billiards are two dimensional enclosures that constrain the motion of
a free particle. Classically, a particle has elastic collisions with
the walls and depending on the geometric properties of the domain, the
dynamics are either regular or chaotic. For the noninteracting problem,
the two particles move independently of each other. In a billiard
system, classical orbits possess simple scaling properties. For
instance, the action of an orbit $\Gamma$, $S_{\Gamma}(\eps) =
\sqrt{2 m \eps} L_{\Gamma}$ and the period of the orbit is
\be\label{Tbill}
T_{\Gamma}(\eps) = {{\partial S_{\Gamma}(\eps)} \over {\partial \eps}}
= {\sqrt{2m} L_\Gamma \over {2 \sqrt{\eps}}}
= {\hbar\sqrt{\alpha}\over 2\sqrt{\eps}}L_\Gamma.
\ee
The parameter $\alpha=2m/\hbar^2$ already appeared in
Eq.~(\ref{pave1pbill}); it will recur often. For example, in all final
expressions, the energy occurs with $\alpha$; this is a result of the
scaling property (the quantity $\alpha E$ having the units of
$1/\mbox{length}^2$.) In the theoretical development, it will be
convenient to retain $\alpha$ and use it to keep track of relative
orders in the semiclassical expansions (since it contains
$\hbar$). However, once we have the final expressions, we are free to
set it to unity for the purposes of numerical comparisons.

We also mention that for billiards, it is common to express the
density of states in terms of the wave number $k$, where
$\eps=k^2/\alpha$ so that $\rho(k) = 2k\rho(\eps)/\alpha$. This is
convenient since $k$ is conjugate to the periodic orbit lengths
$L$. Therefore, many of our results will be quoted as a function of
$k$, although it should be stressed that all convolution integrals
must be done in the energy domain. Thus, we shall write
\be \label{2ppsclk}
\rho_2^{\rm sc}(k) = \left(\bar{\rho}_1*\bar{\rho}_1\right)(k) 
+ 2\left(\bar{\rho}_1*\tilde{\rho}_1\right)(k) +
\left(\tilde{\rho}_1*\tilde{\rho}_1\right)(k).
\ee
Here, it is understood that each of the functions in brackets is first
evaluated in the energy domain and then converted to the $k$ domain
through the Jacobian relation above. This will always be the case when
the argument is $k$, so that we will not always write brackets around
the various functions. In terms of the wavenumber $k$, 
the decomposition (\ref{psaexpandfinE}) becomes
\be\label{symscl}
\rho_{S/A}(k) = {1 \over 2} \left(
\rho_2(k) \pm {1 \over \sqrt{2}} \rho_1\left({k
\over \sqrt{2}} \right) \right).
\ee

\subsection{Smooth Term}

\noindent The smooth part is defined by the convolution integral
\be\label{smthconvint}
\bar{\rho}_1*\bar{\rho}_1(E) = \int_{0}^{E}\d\eps
\bar{\rho}_1(\eps) \bar{\rho}_1(E - \eps),
\ee
where $\bar{\rho}_1$ is given by the Weyl expansion. The expansion in
(\ref{pave1pbill}) is taken only to order $\hbar^0$. Hence, after
expanding the integrand in (\ref{smthconvint}), it is formally
meaningless to include terms that are ${\cal O}\left(1/\hbar\right)$
since there are corrections of the same order in $\hbar$ that have not
been calculated. Ignoring these terms and performing the necessary
integrations, the smooth term is found to be
\be\label{smth2pp}
\bar{\rho}_1*\bar{\rho}_1(E) \approx {\alpha^2
{\cal A}^2 \over 16 \pi^2} E -{\alpha^{3/2}{\cal AL} \over 8 \pi^2}
\sqrt{E} + {\alpha {\cal L}^2 \over 64 \pi}
+ {\alpha{\cal AK} \over 2\pi}.
\ee

\subsection{Cross Term}

\noindent
We next convolve $\bar{\rho}_1$ term by term with $\tilde{\rho}_1$.
Asymptotically, each convolution integral receives contributions from
the upper and lower endpoints. However, we shall only include one of
these, namely the endpoint for which the trace formula is not
evaluated at zero energy. As in section \ref{2ppformalism}, we neglect
the other endpoint for reasons explained in \ref{spurious_explained}
and Ref.\cite{spurious}. This is also discussed in Appendix B, where
we evaluate the various integrals for the cross term exactly using
isolated billiard orbits and show explicitly that an appropriate
asymptotic expansion of the exact expression leads to consistent
results.

After convolution, we find the area term involves the integral
\be \label{int_1}
{\rm Re}\left\{\int_0^E \d\eps A_\Gamma(E-\eps)\exp\left(i\sqrt{\alpha
(E-\eps)}L_\Gamma-i\sigma_\Gamma{\pi\over2}\right)\right\}.
\ee
The lower endpoint $\eps=0$ corresponds to the physically meaningful
situation while the upper endpoint is spurious in the sense mentioned
above and discussed in detail below. Hence, to leading order, we can
remove the amplitude factor from inside the integral, Taylor expand
the argument of the exponential and extend the upper limit to
infinity. This leads to
\be \label{int_1a}
I_{\cal A}(E) \approx {\alpha{\cal A} \over 4\pi^2} \sum_\Gamma {A_\Gamma
\over T_\Gamma} \cos\left(\sqrt{\alpha E}L_\Gamma -
\sigma_\Gamma{\pi\over 2} -{\pi\over 2}\right).
\ee
By similar logic, the perimeter term and curvature terms are
\bea \label{int_23}
I_{\cal L}(E) & \approx & -{\sqrt{\alpha}{\cal L}\over8\pi^{3/2}
\sqrt{\hbar}}\sum_\Gamma {A_\Gamma \over \sqrt{T_\Gamma}}  
\cos\left(\sqrt{\alpha E}L_\Gamma - \sigma_\Gamma{\pi\over 2}
-{\pi\over 4}\right)\nonumber\\
I_{\cal K}(E) & \approx & \phantom{-}{{\cal K} \over \pi\hbar} 
\sum_\Gamma A_\Gamma \cos\left(\sqrt{\alpha E}L_\Gamma -
\sigma_\Gamma{\pi\over 2}\right).
\eea
Note that all amplitudes and periods in (\ref{int_1a}) and
(\ref{int_23}) are evaluated at the system energy $E$. Recall
$\alpha\propto1/\hbar^2$ so that after convolution the sequence is an
expansion in powers of $\sqrt{\hbar}$ and not in powers of $\hbar$ as
for the original Weyl series (\ref{pave1pbill}). We also note that the
first correction to $I_{\cal A}$ may be of the same order as $I_{\cal
K}$ (as happens for the disk \cite{jamal}) and should be included if
this is the case. We then have $\bar{\rho}*\tilde{\rho} \approx
I_{\cal A}+I_{\cal L}+I_{\cal K}$. 

\subsection{Dynamical Term}

\noindent
In this section, we derive a general expression for the dynamical term
that is valid for any billiard problem. To this end, the first task is
to determine the saddle energy from the stationary phase
condition. Inserting (\ref{Tbill}) into (\ref{eqperiods}) yields
\be \label{statcondBnonid}
{L_{\Gamma_1} \over \sqrt{E_0}} =
{L_{\Gamma_2} \over \sqrt{E - E_0}}
\ee
which implies
\be\label{Eo&EmEoiden}
{E_0\over E} = {L_{\Gamma_1}^2 \over L_{\Gamma_1}^2 +
L_{\Gamma_2}^2},
\hspace*{0.5cm} {E - E_0\over E} = {L_{\Gamma_2}^2 \over
L_{\Gamma_1}^2 + L_{\Gamma_2}^2}
\ee
and
\be\label{2ptrfmbillEnonid}
\Upsilon(\Gamma_1,\Gamma_2,E) = -{\sqrt{2m}\over 4E^{3/2}}
{\left({L_{\Gamma_1}^2+L_{\Gamma_2}^2}\right)^{5/2}\over
L_{\Gamma_1}^2L_{\Gamma_2}^2}.
\ee
Clearly $\nu=-1$. We then substitute these results into (\ref{oscosc})
to obtain the two particle trace formula for billiards
\bea \label{ladedah}
\tilde{\rho}_1*\tilde{\rho}_1 (E)  & \approx &
{4E^{3/4}\over \sqrt{\hbar} \alpha^{1/4} (2\pi\hbar)^{3/2}} 
\nonumber\\
&&\sum_{\Gamma_1,\Gamma_2}
{L_{\Gamma_1}L_{\Gamma_2} \over 
\left({L_{\Gamma_1}^2+L_{\Gamma_2}^2}\right)^{5/4}}
A_{\Gamma_1}(E_0)A_{\Gamma_2}(E-E_0)\nonumber\\
&&\cos \left(\sqrt{\alpha E}
\sqrt{L_{\Gamma_1}^2+L_{\Gamma_2}^2} - \left(\sigma_{\Gamma_1} +
\sigma_{\Gamma_2}\right) {\pi \over 2} - {\pi\over 4} \right).
\eea
If the single particle periodic orbits are not
isolated, then one must make direct use of the corresponding single
particle amplitudes in (\ref{trform}) evaluated at the appropriate
energies. We will show an explicit example of this when we analyze the
disk billiard. Note the amplitudes $A_\Gamma$ typically have an energy
dependence so one cannot make any general statements about the
energy dependence of this term except that the greater the
dimensionality of the periodic orbit families, the greater the
energy prefactor. For example, for the disk, it turns out to be
$E^{1/4}$.

If the single particle periodic orbits are isolated, the amplitudes
are given by (\ref{Gutzamp}), which for billiards is 
\be
A_\Gamma(\eps) = {\sqrt{\alpha}\hbar\over2\sqrt{\eps}}
{L_\gamma\over{\sqrt{\left| \det \left( \tilde{M}_{\Gamma} - I \right)
\right|}}}.
\ee
In this case, the Gutzwiller amplitudes are evaluated at $E_0$ and $E -
E_0$, so we again make use of (\ref{Eo&EmEoiden}).  After some
algebra and simplification, we find
\bea\label{2pGutzbillEnonid}   
\tilde{\rho}_1*\tilde{\rho}_1 (E) & \approx & {\alpha^{3/4} \over
(2\pi)^{3/2}E^{1/4}} \nonumber\\ 
&&\sum_{\Gamma_1,\Gamma_2}
{L_{\gamma_1}L_{\gamma_2}(L_{\Gamma_1}^2+L_{\Gamma_2}^2)^{-1/4}
\over 
\sqrt{\left|\det\left(\tilde{M}_{\Gamma_1}-I\right)\right|
      \left|\det\left(\tilde{M}_{\Gamma_2}-I\right)\right|}}
\nonumber\\ 
&&
\cos \left(\sqrt{\alpha E}
\sqrt{L_{\Gamma_1}^2+L_{\Gamma_2}^2} - \left(\sigma_{\Gamma_1} +
\sigma_{\Gamma_2} \right) {\pi \over 2} -{\pi\over 4} \right).
\eea
Note the $E^{-1/4}$ prefactor which implies that the amplitude decays
weakly with energy. This is the same prefactor that occurs in the single
particle disk problem. This is not a coincidence, but arises from the
fact that in both problems the periodic orbits come in one parameter
families. Also, one must be careful to distinguish between $L_\Gamma$,
the length of a periodic orbit and $L_\gamma$, the length of the
corresponding primitive periodic orbit. In general $L_\Gamma=n_\Gamma
L_\gamma$ where $n_\Gamma$ is the repetition index of that orbit.

\subsection{Spurious Endpoint Contributions} \label{spurious_explained}

\noindent
As mentioned above, when confronted with convolution integrals, it is
natural to analyse them asymptotically. This involves identifying the
critical points and doing appropriate expansions in their
neighbourhoods. In our work, these critical points are either
stationary phase points or endpoints. The power of semiclassical
methods is that each critical point can be given an immediate physical
interpretation. For example, the stationary phase point in the
dynamical term is found to be that energy such that the two particles
have the same period so that the motion is periodic in the full
two-particle phase space. This is intuitively reasonable. However, the
same integral also has endpoints with finite valued contributions. We
could do an asymptotic calculation in the vicinity of these points,
but we can argue immediately that the result is spurious and not
physically meaningful.

Recall the trace formulas are asymptotic in $\hbar$ which
typically also means asymptotic in energy. At the endpoints, one of
the trace formulas is evaluated at small energy where it is known to
be invalid. Alternatively, we can substitute for the trace formula
any expression which is asymptotically equivalent to it and
expect all meaningful results to be invariant to leading order. If
we do this, we will find the endpoint contribution changes while
the stationary phase contribution remains invariant, to leading
order.

A further argument is that the structure of the endpoint contribution
will be incorrect. Typically, it will be a sinusoid with an argument
which does not depend on energy, but only depends on the properties of
one of the orbits. Hence, it will have the same asymptotic structure
as the cross term. However, we know that the cross term completely
describes all such contributions and any further contribution with the
same structure must be spurious.

Similarly, when we evaluate the cross term, we have two endpoint
contributions. At one of these, we are evaluating the trace formula at
some finite energy, which is reasonable. This endpoint corresponds to
orbits which are periodic in the full phase space and in which one
particle evolves on a single particle periodic orbit with all the
energy, while the other remains fixed at some point in phase space
with zero energy. At the other endpoint, we are evaluating the trace
formula at zero energy, which is problematic. This corresponds to the
contradictory situation in which the evolving particle has zero energy
while the fixed particle has all the energy. In addition, upon
inspection of this endpoint contribution, we find a function which is
not oscillatory in energy and therefore has the same asymptotic
structure as the smooth term. However, the smooth term already
completely describes the average behaviour of the two particle density
of states and any further contributions with the same structure must
be spurious.

These situations are further examples of a general situation described
in Ref.\cite{spurious} where it was shown that when integrating over
the trace formula to obtain physical quantities, one should include
all critical points except ones at which the trace formula is
evaluated at zero energy. Such contributions should simply be ignored
as spurious. In \cite{spurious}, the application was to thermodynamic
calculations, but the principle is precisely the same. In Appendix B,
we show the result of evaluating the cross term exactly for isolated
orbits. An asymptotic analysis of this result leads to two terms which
we can identify as coming from the two endpoints. One has the form
used in this paper while the other is clearly spurious.

\section{Two Particle Disk Billiard}\label{diskchap}

\noindent In this section, we apply our results to the problem of two
identical noninteracting particles moving in a two dimensional disk
billiard of radius $R$. Quantum mechanically, this
problem is a simple extension of the one body problem. Nevertheless,
the spectrum has some interesting features which we discuss below.

\subsection{Quantum Mechanics}\label{2ppQMcir}

\noindent For the disk billiard, a general two particle state can be
written as  
\be\label{2pstatecir}
   \left | m_1 \hspace*{0.1cm} n_1 , m_2 \hspace*{0.1cm} n_2 \right
\rangle = \left | m_1 \hspace*{0.1cm} n_1 \right \rangle \otimes \left
| m_2 \hspace*{0.1cm} n_2 \right \rangle 
\ee
where the azimuthal quantum numbers $m_1$, $m_2=0, \pm 1, \pm 2,
\ldots$ and the radial quantum numbers $n_1$, $n_2=1, 2, 3,
\ldots$. We shall also use a more compact notation
$\left | N_1 , N_2 \right \rangle = \left | m_1 \hspace*{0.1cm} n_1 ,
m_2 \hspace*{0.1cm} n_2 \right \rangle$ where $N$ denotes a pair of
integers ($m,n$). We can immediately write down the wave numbers of
the two particle system as
\be\label{2pkscir}
   k_{N_1 N_2} = \sqrt{\left({Z_{N_1} \over R}\right)^2 +
\left({Z_{N_2} \over R}\right)^2}, 
\ee
where $Z_N$ denotes the $n$th zero of the $m$th Bessel function
$J_m(z)$.  The set of all two particle states is given by
$\left\{\left|N_1,N_2\right\rangle\right\}$.

The spectrum is highly degenerate. A typical state $\left| N_1 , N_2
\right \rangle$ is 8-fold degenerate since we can reverse the sign of
either $m_1$ or $m_2$ or interchange the two particles and the
resultant state has the same energy. However, if either $m_1$ or $m_2$
is zero or if $N_1=N_2$ then the state is 4-fold degenerate.  If
$m_1=m_2=0$ and $N_1 \neq N_2$, then the state is 2-fold degenerate
whereas if $m_1=m_2=0$ and $N_1=N_2$, then the state is nondegenerate.
If the particles are in distinct states, the degenerate multiplets
divide evenly between the symmetric and antisymmetric spaces. However,
if the particles are in the same state, $N_1=N_2$, it is somewhat less
trivial. If $N_1=N_2$ and $m_1=m_2\neq 0$, there is a 4-fold
degenerate set of states: $\left | m \hspace*{0.1cm} n , m
\hspace*{0.1cm} n \right \rangle$, $\left | -m \hspace*{0.1cm} n , -m
\hspace*{0.1cm} n \right \rangle$, $\left | m \hspace*{0.1cm} n
, -m \hspace*{0.1cm} n \right \rangle$ and $\left | -m
\hspace*{0.1cm} n , m \hspace*{0.1cm} n \right \rangle$. The first two
states belong to the symmetric space. From the second two states, we
can construct one symmetric and one antisymmetric combination. (This
is analogous to coupling two spin 1/2 states to construct a 3-fold
symmetric $S=1$ state and a nondegenerate antisymmetric $S=0$ state.)
If $N_1=N_2$ and $m_1=m_2=0$, this yields the state $\left |0
\hspace*{0.1cm} n , 0 \hspace*{0.1cm} n \right \rangle$, which is
singly degenerate and belongs to the symmetric space.

\begin{figure}[h]
\vspace*{-3.5cm}\hspace*{-1.1cm} 
\psfig{figure=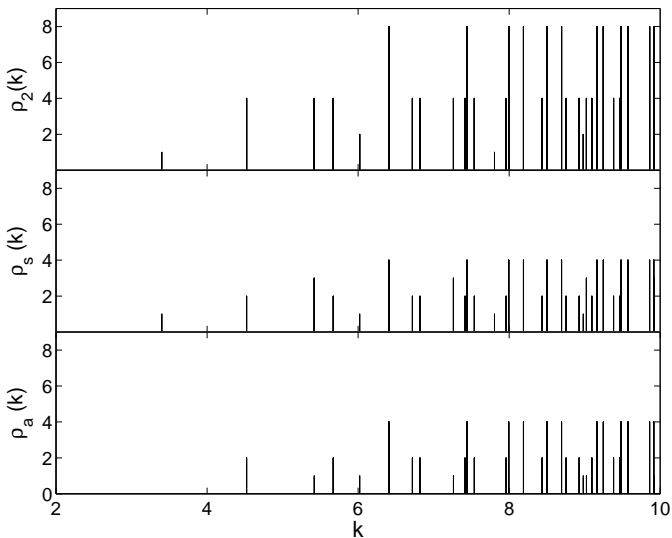,height=5.5in}
\vspace*{-3.0cm}
\caption[]{\small (Top) The quantum density of states for two
identical particles in the disk billiard. (Middle) Bosonic density of
states. (Bottom) Fermionic density of states. In each case, the
heights indicate the degeneracy of the state.}
\label{diskqtm}
\end{figure}

The quantum density of states
\be\label{pqmfull}
    \rho_2(k) = \sum_{N_1,N_2} \delta(k-k_{N_1 N_2})
\ee
and the corresponding symmetric and antisymmetric densities are shown
in Fig.~\ref{diskqtm} as a function of the wavenumber $k$. Note that 
in this figure some of the peaks have different degeneracies in the
symmetric and antisymmetric densities, as discussed above.

\subsection{Semiclassical Density of States}

\noindent We first review the semiclassical decomposition of the
single particle density of states. The smooth part of the density of
states may be obtained using the general result for two dimensional
billiards (\ref{pave1pbill}). In fact, many higher order terms have
been calculated \cite{howls}. But, for our purposes, it suffices to
use the first three terms as in (\ref{pave1pbill}) with ${\cal A} =
\pi R^2$, ${\cal L}=2\pi R$ and ${\cal K} = 1/6$.

The oscillating part of the level density can be obtained using trace
formulas for systems with degenerate families of orbits. The periodic
orbit families may be uniquely labelled by two integers (${\rm v}$,
${\rm w}$) where ${\rm v}$ is the number of vertices and ${\rm w}$ is
the winding number around the center. The two integers must satisfy
the relation ${\rm v} \ge 2 {\rm w}$. The length of an orbit with
vertex number ${\rm v}$ and winding number ${\rm w}$ is given by
$L_{\rm vw}=2{\rm v}R\sin\left(\pi{\rm w}/{\rm v}\right)$. With this
notation, the trace formula for the oscillating part of the density of
states is \cite{Reimann}
\bea\label{posc1pcirE}
\tilde{\rho}_1(\eps) & \approx & {\alpha^{3/4} \over2\sqrt{2\pi}\eps^{1/4}}
\sum_{\rm vw}
{{\cal D}_{\rm vw} L_{\rm vw}^{3/2}\over {\rm v}^2}\nonumber\\
&& \cos \left( \sqrt{\alpha E} L_{\rm vw} - 3{\rm v}{\pi \over
2} + {\pi \over 4}\right)
\eea
where the sum goes from ${\rm w}=1\cdots\infty$ and ${\rm v}= 2{\rm
w}\cdots\infty$ and the degeneracy factor ${\cal D}_{{\rm v}{\rm w}}$,
which accounts for negative windings, is $1$ for ${\rm v} = 2 {\rm w}$
and $2$ for ${\rm v} > 2 {\rm w}$. Comparing (\ref{posc1pcirE}) with
the general form (\ref{trform}), we identify
\bea\label{diskamp}
A_{\rm vw}(\eps) & = &{\sqrt{2\pi}\alpha^{3/4}\hbar
{\cal D}_{\rm vw} L_{\rm vw}^{3/2}\over 4{\rm
v}^2\eps^{1/4}}\nonumber\\
\sigma_{\rm vw} & = & 3{\rm v} - {1\over2}
\eea
Adding the smooth and oscillating terms gives the semiclassical
approximation to the single particle density of states which we denote
by $\rho_1^{\rm sc}(\eps)$.

To evaluate the semiclassical approximation to the two particle
density of states, we must evaluate the smooth, cross and dynamical
terms. The smooth term can be taken from Eq.~(\ref{smth2pp}) to be
\be\label{2pweylE}
\bar{\rho}_1*\bar{\rho}_1(E) \approx
{\alpha^2 R^4\over 16}E - {\alpha^{3/2}R^3\over4}\sqrt{E}
+\left({3\pi+4\over48}\right)\alpha R^2.
\ee
The arguments of the previous section and in particular
Eqs.~(\ref{int_1a}) and (\ref{int_23}) lead to the cross
term
\bea\label{2pcrossE}
&&\bar{\rho}_1*\tilde{\rho}_1 (E) \approx
{\alpha^{5/4}R^2E^{1/4}\over4\sqrt{2\pi}}
\sum_{\rm vw}{\sqrt{L_{\rm v w}}{\cal D}_{\rm vw}\over {\rm
v}^2}\left(\cos\left(\Phi_{\rm
vw}-{\pi\over2}\right)\right. \nonumber\\ 
&&\left.- \sqrt{\pi\over2}\chi_{\rm vw}\cos\left(\Phi_{\rm vw}-
{\pi\over4}\right)
+\left({1\over3}+{R^2\over2L_{\rm vw}^2}\right)\chi_{\rm
vw}^2\cos\Phi_{\rm vw}\right)
\eea
where $\Phi_{\rm vw}=\sqrt{\alpha E}L_{\rm vw}-3{\rm v}\pi/2+\pi/4$
and $\chi_{\rm vw}=\sqrt{L_{\rm vw}}/(\alpha E)^{1/4}R$. We have also
included the first correction to the area term, $I_{\cal A}(E)$ which
appears in the third term above \cite{jamal}. The dynamical term can
be obtained using (\ref{ladedah}). Noting that $\Gamma_i$ in
(\ref{ladedah}) corresponds to the pair of integers (${\rm v}_i$,${\rm
w}_i$), the result is
\bea\label{2poscE}
\tilde{\rho}_1*\tilde{\rho}_1(E) & \approx &
{\alpha^{5/4}E^{1/4}\over4\sqrt{2\pi}}
\sum_{\rm v_1 w_1 , v_2 w_2}
\left(\prod^2_{i=1} {{\cal D}_iL_i^2\over{\rm v}_i^2}\right)
L_{12}^{-3/2}\nonumber\\ 
&&\cos \left(\sqrt{\alpha E}L_{12} - 3 \left({\rm v}_1 + {\rm v}_2
\right){\pi\over2} + {\pi\over4}\right)
\eea
where we defined $L_i=L_{{\rm v}_i}{{\rm w}_i}$, ${\cal D}_i={\cal
D}_{{\rm v}_i}{{\rm w}_i}$ and $L_{12}=\sqrt{L_1^2 + L_2^2}$.

\begin{figure}[h]
\vspace*{-3.5cm}\hspace*{-1.1cm} 
\psfig{figure=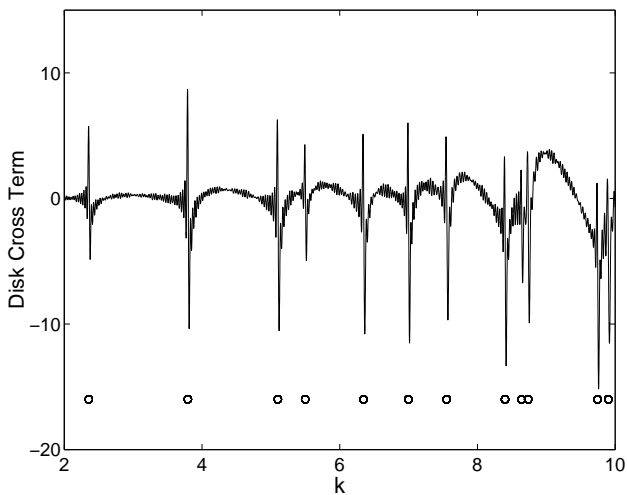,height=5.5in}
\vspace*{-3.0cm}
\caption[]{The cross term (\ref{2pcrossE}) of the semiclassical
density of states for two identical particles in the disk billiard. In
this case, we truncate the sum in (\ref{2pcrossE}) at ${\rm
w_{max}}=50, {\rm v_{max}}=100$.  The circles indicate the level
sequence of the one body problem obtained from EBK quantization. Note
the kinks that occur at these positions.}\label{2pcross}
\end{figure}

\subsection{Numerics}

\begin{figure}[h]
\vspace*{-3.5cm}\hspace*{-1.1cm} 
\psfig{figure=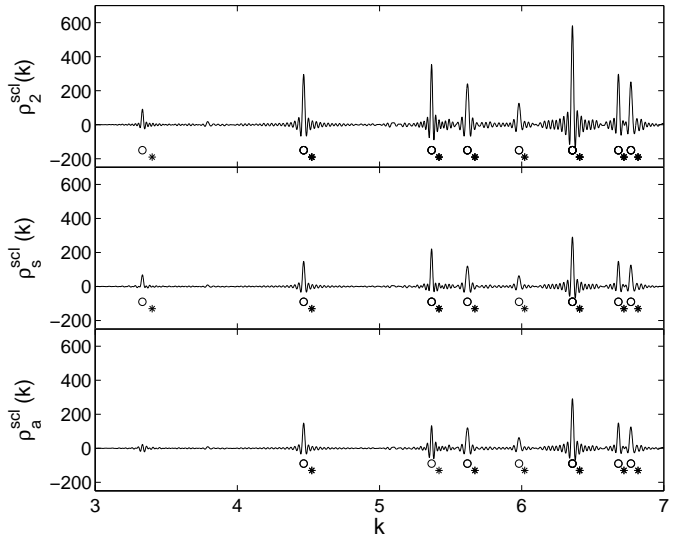,height=5.5in}
\vspace*{-3.0cm}
\caption[]{(Top) The first few peaks of the semiclassical density of
states (\ref{2ppsclk}). (Middle) Semiclassical approximation to the
bosonic density of states. (Bottom) Semiclassical approximation to the
fermionic density of states. In each case, the circles and stars
represent the appropriate level sequences obtained from EBK
quantization and quantum mechanics, respectively. Note the positions
of the peaks more closely reproduce the EBK spectrum.}\label{diskraw}
\end{figure}

\begin{figure}[h]
\vspace*{-3.5cm}\hspace*{-1.1cm} 
\psfig{figure=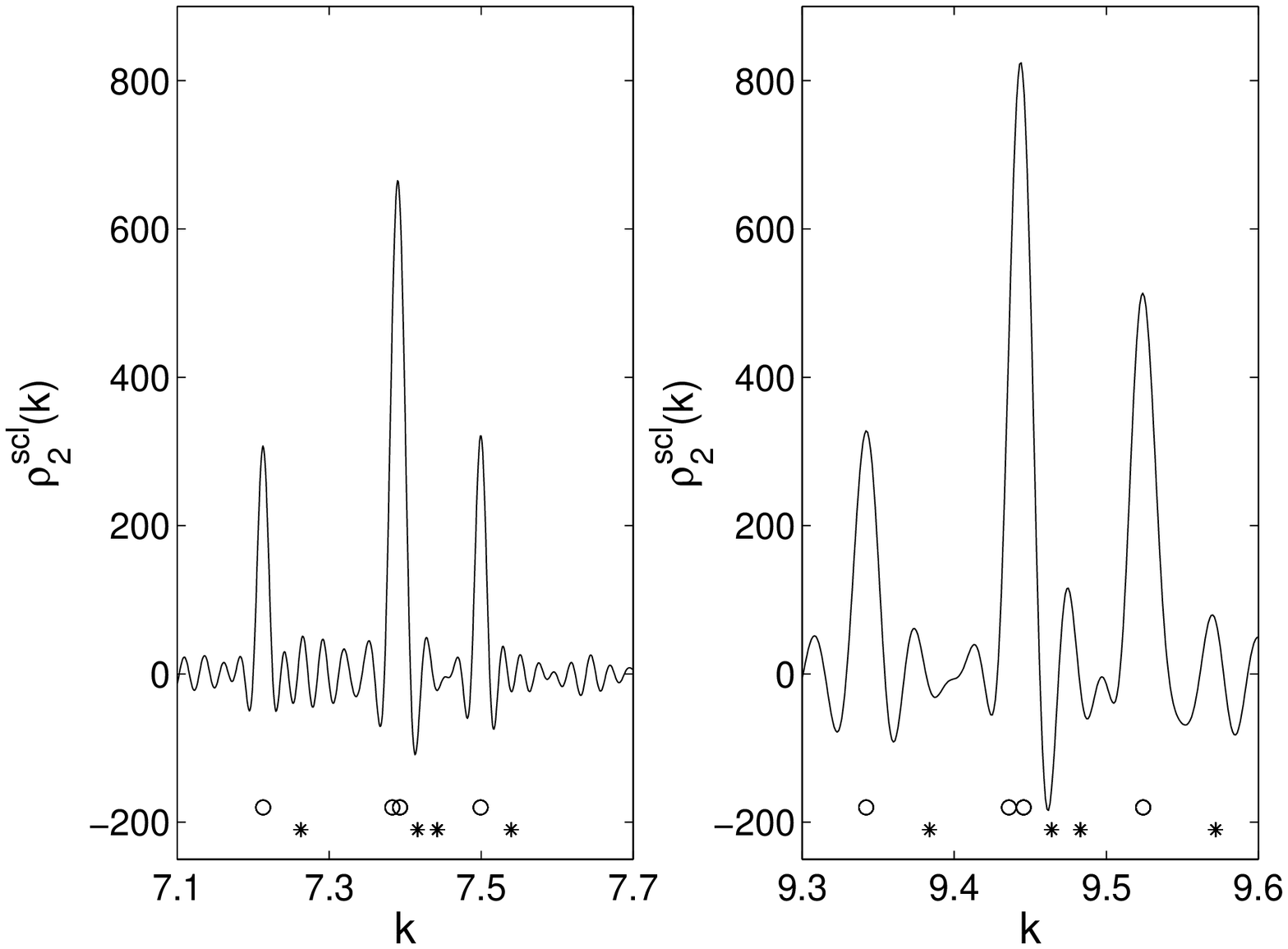,height=5.5in}
\vspace*{-3.0cm}
\caption[]{Two sets of two peaks in the semiclassical density of
states (\ref{2ppsclk}) that are not resolved. (Left) The middle peak
is not resolved into the two peaks at
$k=7.4163$ (corresponding to the
quartet $\left\{ \left| 0 \hspace*{0.1cm} 1 , \pm 1 \hspace*{0.1cm} 2
\right \rangle, \left| \pm 1 \hspace*{0.1cm} 2 , 0 \hspace*{0.1cm} 1
\right \rangle \right\}$) and $k=7.4423$ (corresponding to
the octet $\left\{ \left| \pm 1 \hspace*{0.1cm} 1 , \pm 3
\hspace*{0.1cm} 1  \right \rangle, \left| \pm 3 \hspace*{0.1cm} 1 ,
\pm 1 \hspace*{0.1cm} 1  \right \rangle \right\}$). The corresponding
EBK quartet and octet energies occur at $k=7.3831$ and $k=7.3932$
respectively. Note that $\Delta k_{EBK} = 0.0101$ and $\Delta k_{QM} =
0.026$ so that the spacing of the two unresolved levels is smaller in
the semiclassical spectrum than in the quantum mechanical spectrum.
(Right) The middle peak is not resolved into the two peaks at
$k=9.4641$ (corresponding to the quartet $\left\{ \left| \pm 1
\hspace*{0.1cm} 1 , 0 \hspace*{0.1cm} 3 \right \rangle, \left| 0 
\hspace*{0.1cm} 3 , \pm 1 \hspace*{0.1cm} 1  \right \rangle \right\}$)
and $k=9.4829$ (corresponding to the octet $\left\{\left| \pm 3 
\hspace*{0.1cm} 1 ,  \pm 1 \hspace*{0.1cm} 2  \right 
\rangle,\left| \pm 1 \hspace*{0.1cm} 2 , \pm 3 \hspace*{0.1cm} 1
\right \rangle \right\}$.) The corresponding EBK quartet and octet
energies occur at $k=9.4359$ and $k=9.4456$ respectively. Here,
$\Delta k_{EBK} = 0.0097$ and $\Delta k_{QM} = 0.0188$.}
\label{2pnores}
\end{figure}

\noindent 
For numerical purposes, we take $\alpha=\hbar=1$ and $R=1$ so the
single-particle energies are just the squares of the zeros of Bessel
functions. Since we can only include a finite number of orbits, the
periodic orbit sums must be truncated. As a representative case, we
truncate the sum in (\ref{2pcrossE}) at ${\rm w}_{\rm max}=50, {\rm
v}_{\rm max}=100$ (see Fig.~\ref{2pcross}) and use the same limits to
truncate the quadruple sum in (\ref{2poscE}). This is a relatively
small set of orbits, yet it does very well in reproducing the peaks of
the quantum density of states. As an illustration, we show the first
few peaks of (\ref{2ppsclk}) in Fig.~\ref{diskraw}. We calculated the
semiclassical density of states (\ref{2ppsclk}) on the interval $0 \le
k \le 11$. After doing so, we found only two sets of two peaks which
were not resolved. These are shown in Fig.~\ref{2pnores}. Obviously,
using more orbits will produce better results, but this increases the
computation time because of the quadruple sum in (\ref{2poscE}).
(Although one can reduce the computational overhead by limiting the
sum to orbits whose amplitude exceeds some prescribed threshold
\cite{jamal}).

As an additional test, we want to determine whether (\ref{2ppsclk})
gives the correct degeneracies. We could do this by integrating the
area under each of the peaks. However, a simpler procedure is to do a
Gaussian smoothing by convolving $\rho_2^{\rm sc}(k)$ with an
unnormalized Gaussian of variance $\sigma$:
\be\label{2pcirsclsmt}
\rho_2^{\rm sc}(k) *
G_{\sigma}(k) = \int_{0}^{\infty}\d k' \rho_2^{\rm sc}(k')
G_{\sigma}(k - k')
\ee
where 
\be\label{Gsmth}
G_{\sigma}(k) = \exp(-k^2/{2 \sigma^2}).
\ee
and $\sigma$ is the smoothing width. The reason for this is that if
the variance $\sigma$ of the Gaussian is larger than the intrinsic
width of a peak in the semiclassical spectrum, then each peak acts
like $d \delta(k-k_n)$ with respect to the Gaussian. Thus, the
integral in (\ref{2pcirsclsmt}) becomes $d G_{\sigma}(k-k_n)$ or $d$
at $k=k_n$. Of course, this is invalid when the spacing between two
adjacent peaks is smaller than about $\sigma$. Some examples are
discussed in the next section.

\begin{figure}[h]
\vspace*{-3.5cm}\hspace*{-1.1cm} 
\psfig{figure=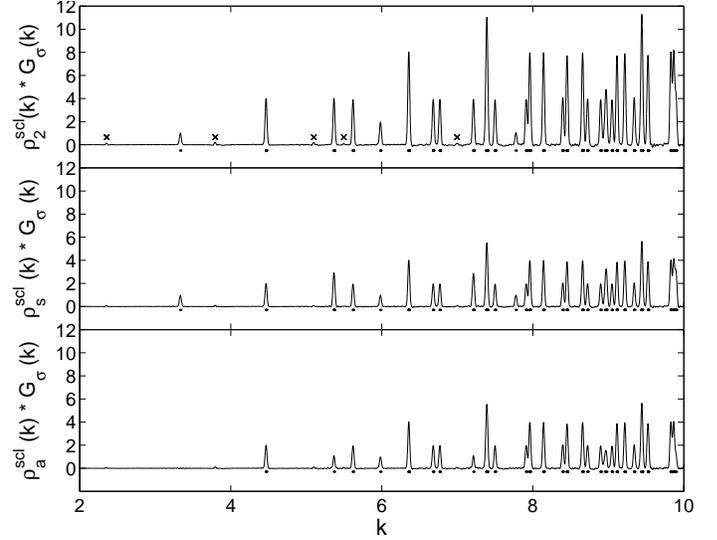,height=5.5in}
\vspace*{-3.0cm}
\caption[]{(Top) The smoothed semiclassical density of states obtained
from numerical convolution of (\ref{2ppsclk}) with (\ref{Gsmth}).
Note the artifacts of the single particle EBK spectrum which occur at
the positions marked by an ``X''. (Middle) The smoothed semiclassical
bosonic density of states obtained from numerical convolution of
(\ref{symscl}) (with the + sign) and (\ref{Gsmth}). (Bottom) The
smoothed semiclassical fermionic density of states obtained from
numerical convolution of (\ref{symscl}) (with the - sign) and
(\ref{Gsmth}). In each case the sequence of dots represent the
corresponding EBK spectrum and $\sigma=0.0125$.}\label{disksmooth}
\end{figure}

We also studied the symmetrised densities by using the expression
(\ref{symscl}) for both the quantum and semiclassical densities and
convolving as above. The periodic orbit sums in the oscillating parts
of the one and two body densities were truncated in the standard
manner as before. The result of this numerical procedure is shown in
Fig.~\ref{disksmooth}. Clearly, the semiclassical approximations
reproduce the correct degeneracies of the quantum spectrum as well as
the approximate positions.

\subsection{Discussion}

\noindent 
In \cite{Reimann}, it was noted that the trace formula replicates the
single particle EBK spectrum obtained from torus quantisation more
precisely than it duplicates the exact single particle quantum
spectrum. After inspection of Figs.~\ref{diskraw} and
\ref{disksmooth}, we notice the same effect in the two particle
spectrum. This property of the trace formulas also accounts for the
unresolved peaks in the semiclassical spectrum. When the spacing of
two levels of the EBK spectrum is very small, our truncated trace
formulas may not resolve them, regardless of the spacing of the
corresponding levels in the quantum spectrum (cf. Fig.\ref{2pnores}).

Comparing Figs.~\ref{diskqtm} and \ref{disksmooth}, we observe
generally good agreement between the quantum and semiclassical
spectra. Still, there are some apparent inconsistencies, for example,
the two tall peaks in Fig.~\ref{disksmooth}. These are the two sets of
unresolved levels in Fig.~\ref{2pnores}, in each case an octet and a
quartet. The reason for the discrepancy is the level spacings are
smaller than the smoothing width $\sigma$, in contradiction to the
assumption above, so that the peak height does not equal the
degeneracy.  In fact, the peak heights observed are rather close to 12
since the octets and quartets are very nearly degenerate on the scale
of $\sigma$ and act almost like a 12-fold degenerate set. It is not
perfectly $12$ due to the fact that the degeneracy is not perfect.
However, we also observe that the integrated weight under the peak is
consistent with a set of 12 energy levels. Other inconsistencies in
Fig.~\ref{disksmooth} occur for the same reason. Of course, overall
improvements can be made by including more orbits.  As well, the
artifacts of the single particle spectrum (some examples are marked by
an ``X'' in Fig.~\ref{disksmooth}) which arise from errors in the
cross term, presumably decrease when corrections to the single
particle trace formula are incorporated into the cross term. These
preliminary numerical findings support our analytical results, which we
now test in the rather different context of a chaotic billiard.

\section{Two Particle Cardioid Billiard}\label{2ppcard}

\noindent
In this section, we study the problem of two identical noninteracting
particles evolving in the cardioid billiard, which is fully chaotic
\cite{Markarian}. Since the billiard has a reflection symmetry, all
the quantum states are either even or odd (this symmetry should not be
confused with the symmetric/antisymmetric symmetry due to particle
exchange.) In the subsequent analysis, we will exclusively use the odd
spectrum. The reason for this is to avoid the additional complication
of {\em diffractive} orbits which strike the vertex. Classically,
these orbits are undefined and are therefore not included in the
standard Gutzwiller theory. Studies of diffractive effects in trace
formulas can be found in \cite{diff,Bruus}. The latter reference
explores the specific application to the cardioid and shows that
diffractive orbits are important in describing the even spectrum but
are largely absent from the odd spectrum.

\begin{figure}[h]
\vspace*{-4.5cm}\hspace*{-1.1cm} 
\psfig{figure=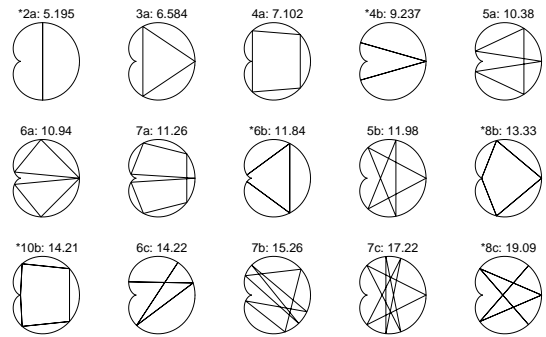,height=5.5in}
\vspace*{-4.0cm}
\caption[]{Some of the shorter periodic orbits of the cardioid
in the full domain. The label of each orbit includes the number of
reflections and also a letter index to further distinguish it. The two
orbits *8b and *10b reflect specularly near the cusp, contrary to
appearances while the orbit 4a misses the cusp. From
\cite{Bruus}.}\label{Bruusfig}
\end{figure}

We could proceed as before by doing an explicit semiclassical analysis
of each term in the decomposition of the two particle semiclassical
density of states (\ref{2pdecomp}). However, we can simplify the
analysis by removing single particle dynamics from the discussion.
That is, we will focus exclusively on those quantum mechanical and
semiclassical quantities that inherently describe two particle
dynamics. More specifically, we compare the Fourier transform of the
dynamical term
\be\label{smcl2pptrans}
\tilde{F}_2^{\rm sc}(L)={\cal F} \{ \tilde{\rho}_1 * \tilde{\rho}_1(k) \}
\ee
with its quantum mechanical analogue which we define to be 
\be\label{qm2pptrans}
\tilde{F}_2^{\rm qm}(L)={\cal F} \{\rho_2(k) - \bar{\rho}_1*
\bar{\rho}_1(k) - 2\bar{\rho}_1*\tilde{\rho}_1(k) \}.  
\ee
The integral operator ${\cal F}$ will be defined precisely below.  In
the semiclassical transform (\ref{smcl2pptrans}), we use
(\ref{2pGutzbillEnonid}) expressed in terms of the wavenumber. Here,
$\Gamma_1$ and $\Gamma_2$ are periodic orbits in the fundamental
domain ({\it i.e.} the half-cardioid.) Orbit properties are discussed
in \cite{Backer,Bruus} and some representative orbits are shown in
Fig.~\ref{Bruusfig}. The stability matrices in the denominators of the
single particle Gutzwiller amplitudes are computed using the standard
prescription for the stability of free flight billiards (see, for
example, \cite{ChaosBook}.)

In the quantum mechanical analogue (\ref{qm2pptrans}), $\rho_2(k)$ is
the quantum two particle density of states $\rho_2(k) = \sum_{I}
\delta \left(k - k_I \right)$ where the superindex $I$ denotes the
pair of integers ($i,j$) and $k_I = \sqrt{k_i^2 + k_j^2}$. In
(\ref{qm2pptrans}), we subtract the smooth average part and the part
which contains single particle dynamics. Using ${\cal
A}=3\pi/4$, ${\cal L}=6$, and ${\cal K}=3/16$ in
(\ref{smth2pp}), the smooth term is
\be\label{2pweylcardE}
\bar{\rho}_1 * \bar{\rho}_1(E) \approx {9 \over 256}\alpha^2 E
- {9 \over 16\pi}\alpha^{3/2}\sqrt{E} +
\left({9 \over 16\pi} +  {9 \over 128}\right)\alpha.
\ee
The cross term is given by the general expressions (\ref{int_1a}) and
(\ref{int_23}).

\subsection{Numerics for the Unsymmetrised Cardioid}\label{2ppcardnum}

\noindent  As before, we take $\alpha=1$ and use a standard
sized cardioid as in \cite{Bruus} to obtain the single particle
spectrum. In this section, we numerically compare the two particle
quantum mechanics with the two particle semiclassics. We do this by
making a direct comparison of the Fourier transforms in the reciprocal
space of orbit lengths, $L$. In this space, we expect peaks at lengths
which correspond to the Euclidean lengths of the {\em full} periodic
orbits of the two particle system. For instance, if the full orbit
$\Gamma$ is comprised of particle 1 travelling on the orbit $\Gamma_1$
and particle 2 traversing a distinct orbit $\Gamma_2$, we expect a
peak at $L_{\Gamma}=\sqrt{L^2_{\Gamma_1} + L^2_{\Gamma_2}}$. In the
event that both particles are on the same orbit $\Gamma$, we expect a
peak at $\sqrt{2}L_{\Gamma}$. In this way, any peak in the two
particle spectrum can be attributed to the dynamics of a particular
periodic orbit of the full classical phase space.

We construct the two particle spectrum by adding the energies of the
single particle spectrum. We include the first 1250 single particle
energies which allows us to construct the first $766794$ two particle
energy levels representing all two particle energies less than $6.8856
\times 10^3$.

\begin{figure}[h]
\vspace*{-3.5cm}\hspace*{-1.1cm} 
\psfig{figure=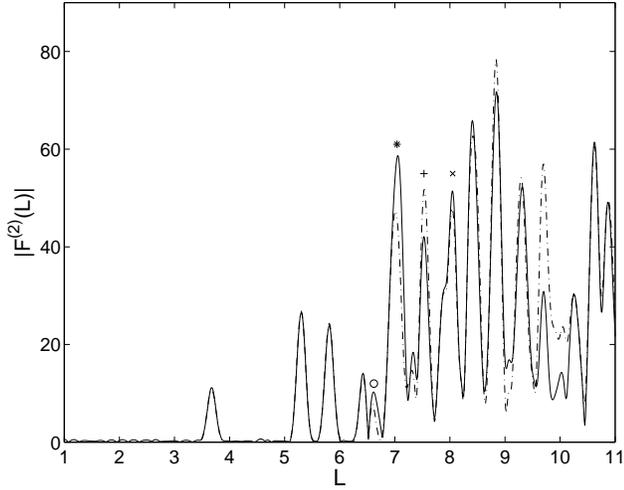,height=5.5in}
\vspace*{-3.0cm}
\caption[]{The Fourier transform of the dynamical (purely oscillating)
part of the two particle density of states. The solid line is the
transform of the quantum mechanical two particle spectrum
(\ref{qm2pptrans}) and the dashed-dotted line is the transform of the
semiclassical two particle trace formula (\ref{smcl2pptrans}). All
relevant geometrical periodic orbits with length
$L_{\Gamma}=\sqrt{L^2_{\Gamma_1} + L^2_{\Gamma_2}} < 11$ have been
included.  (Symbols described in the text.)}\label{cardtot2}
\end{figure}

For a precise numerical comparison, we define the Fourier transform
\be\label{Fourier}
{\cal F}\{f(k)\} = \int_{-\infty}^{\infty}\d k {\rm w}(k) e^{i kL}
f(k)
\ee
as a function of the conjugate variable $L$. Here, ${\rm w}(k)$
is the three term Blackman-Harris window function
\cite{Harris}
\be\label{BHW1}
{\rm w}(k) = \left\{\begin{array}{lrl} \sum_{j=0}^{2} a_j \cos \left(
{2 \pi j} {{k-k_0} \over {k_f-k_0}} \right) && k_0 < k < k_f \\  
0 && \mbox{otherwise}
\end{array} \right.
\ee
with $(a_0,a_1,a_2) = (0.42323,-0.49755,0.07922)$. We choose $k_0$ and
$k_f$ so that the window function goes smoothly to zero at the first
and last eigenvalues of the two particle spectrum.  Numerical
integration of (\ref{smcl2pptrans}) and (\ref{qm2pptrans}) using this
definition of ${\cal F}$ is displayed in Fig.~\ref{cardtot2}. In the
semiclassical transform, a total of 100 periodic orbits including
multiple repetitions were used.

In Fig.~\ref{cardtot2}, we observe good agreement between the quantum
and semiclassical results for $L<6.5 $ and $L>10.3$. In the region
$6.5 < L < 10.3$, there are appreciable discrepancies for the
following reason. Recall that the amplitudes of the two particle trace
formula (\ref{2pGutzbillEnonid}) apply only to billiard systems whose
single particle periodic orbits are isolated. In the single particle
cardioid problem, there exist orbits which are not well isolated in
phase space, in fact two geometric orbits and a diffractive orbit are
sometimes very close in phase space. For example, the two geometric
orbits $4a$ and $^*10b$ together with the similar looking diffractive
orbit $4a'$ (not shown) \cite{Bruus}. In this event, the stationary
phase approximation underlying the Gutzwiller formalism fails as does
the argument in that diffractive orbits do not affect the odd
spectrum. As a result, whenever a two particle orbit in the full space
is comprised of one or both particles on one of these problematic
single particle periodic orbits, the resulting two particle amplitude
is inaccurate. (There is recent work on uniform approximations to
account for such effects \cite{Sieber}, unfortunately it seems not to
apply to the cardioid which has the additional curious feature that
the boundary curvature is infinite at the vertex.)

\begin{table} \label{table1}
\begin{tabular}{cccc}
${\Lambda}$ & ${\Gamma_1}$ & ${\Gamma_2}$ & ${L_{\Lambda}}$ \\
        $1$ & ${1\over2}(^*2{\rm a})$ & $4{\rm a}$ & $7.562$ \\
        $2$ & ${1\over2}(^*2{\rm a})$ &${1\over2}(^*10{\rm b})$ & $7.565$
\end{tabular}
\caption{A few of the periodic orbits which conspire to give
trouble around L=7.6.}
\end{table}

\begin{figure}[h]
\vspace*{-3.5cm}\hspace*{-1.1cm} 
\psfig{figure=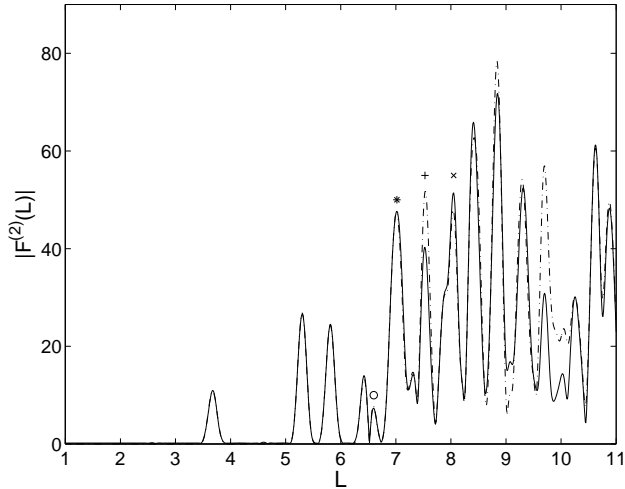,height=5.5in}
\vspace*{-3.0cm}
\caption[]{Same as figure \ref{cardtot2} except that the cross term
in (\ref{qm2pptrans}) is computed using single particle quantum
mechanics. (Symbols described in the text.)}\label{cardtot1}
\end{figure}

\begin{figure}[h]
\vspace*{-3.5cm}\hspace*{-1.1cm} 
\psfig{figure=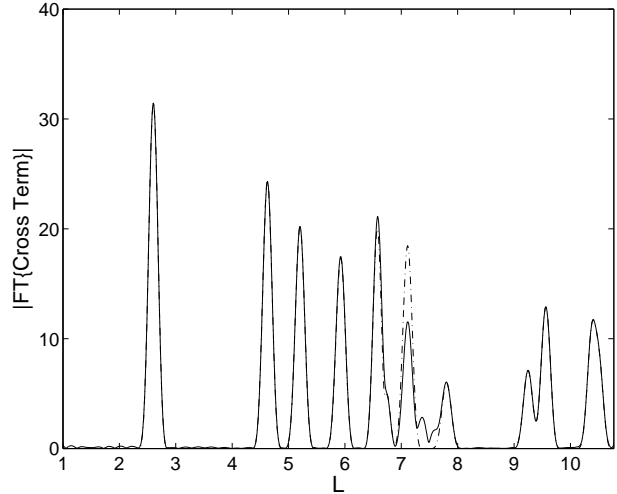,height=5.5in}
\vspace*{-3.0cm}
\caption[]{The cross term for the cardioid billiard calculated using
quantum mechanics (solid) and periodic orbit theory
(dashed-dotted).}\label{cardcross} 
\end{figure}

We now consider some specific examples. Consider first the peak
structures ``o'' and ``*''. In this region, the single particle trace
formula is erroneous \cite{Bruus} and these errors propagate through
to the cross term and inevitably to the quantum mechanical
transform. We have also computed the cross term using quantum
mechanics, that is, using $\tilde{\rho}_1 = \rho_1-\bar{\rho}_1$ in
(\ref{qm2pptrans}) and confirmed that these discrepancies do not arise
(cf. Figs.~\ref{cardtot1} and \ref{cardcross}). Thus, these
discrepancies are due to errors in the semiclassical approximations of
the cross term.

For the rest of the discussion, $\Lambda$ refers to a particular
periodic orbit family in the full phase space with each two particle
orbit in this family comprised of the single particle orbits
$\Gamma_1$ and $\Gamma_2$ and $L_{\Lambda}$ are the lengths of the
orbits in each family.  Next, consider the peak structure at $L
\approx 7.5$ (+). There are two families of orbits, $\Lambda_1$ and
$\Lambda_2$ that are responsible for these peaks. The underlying
structures of these orbits are shown in Table~I. Bearing in mind the
two single particle orbits $\Gamma_2$ are not well isolated
(cf. Fig.~\ref{Bruusfig}), the Gutzwiller amplitude of each $\Gamma_2$
is incorrect. Consequently, the two particle Gutzwiller amplitude will
also be incorrect, as Fig.~\ref{cardtot2} demonstrates.  Let us look
at the next peak structure. Clearly, the quantum peak heights are
underestimated at $L \approx 8.1$ (X). We account for this by
recognizing the two particle orbit structure involves the single
particle orbit ${1\over2}(^*8{\rm b})$ which is an orbit which passes
close to the vertex. More specifically, the orbit family ${\Lambda}=3$
is composed of single particle orbits ${\Gamma_1}={1\over2}(^*4{\rm
b})$ and ${\Gamma_2}={1\over2}(^*8{\rm b})$ and the lengths of these
two particle orbits are ${L_{\Lambda}}= 8.109$. As a final
illustration, we consider the region $9.6 < L < 10.3$. In this
neighbourhood, the semiclassics are particularly bad. This can be
accounted for by inspection Table~II.

As the Table~II and Fig.~\ref{Bruusfig} show, there are many instances
where both of the single particle orbits constituting the full orbit
are poorly isolated. In view of this, both single particle Gutzwiller
amplitudes are incorrect making the product even worse.  This accounts
for the gross inconsistencies in this region of the reciprocal
space. The other discrepancies can be accounted for in a similar
manner.

\begin{table} \label{table2}
\begin{tabular}{cccc}
${\Lambda}$ & ${\Gamma_1}$ & ${\Gamma_2}$ & ${L_{\Lambda}}$ \\
$4$ & $3{\rm a}$ & $4{\rm a}$ & $9.684$ \\
$5$ & $3{\rm a}$ & ${1\over2}(^*10{\rm b})$ & $9.687$ \\
$6$ & $4{\rm a}$ & ${1\over2}(^*8{\rm b})$  & $9.740$ \\
$7$ & ${1\over2}(^*8{\rm b})$ & ${1\over2}(^*10{\rm b})$ & $9.742$ \\
$8$ & $4{\rm a}$ & $4{\rm a}$ & $10.044$ \\
$9$ & $4{\rm a}$ & ${1\over2}(^*10{\rm b})$ & $10.046$ \\
$10$ & ${1\over2}(^*10{\rm b})$ & ${1\over2}(^*10{\rm b})$ & $10.048$ 
\end{tabular}
\caption{A few of the periodic orbits which conspire to give
trouble around L=10.}
\end{table}

\subsection{Symmetry Decomposition}

\noindent
In this section, we explore the symmetry decomposition of the two
particle problem and in particular the comparison of the symmetrized
two particle quantum mechanics with the corresponding two particle
semiclassics. We start by defining the smooth and oscillating symmetry 
reduced densities of states from (\ref{symscl})
\be\label{PSAsmooth}
\bar{\rho}_{S/A}(k)={1 \over 2}\left(
(\bar{\rho}_1*\bar{\rho}_1)(k) \pm {1
\over \sqrt{2}} \bar{\rho}_1 \left ({k \over \sqrt{2}} \right) \right)
\ee
and
\be\label{PSAosc}
\tilde{\rho}^{\rm dyn}_{S/A}(k) =
{1\over 2} \left((\tilde{\rho}_1*\tilde{\rho}_1)(k)
\pm  {1 \over \sqrt{2}} \tilde{\rho}_1\left({k \over \sqrt{2}}\right)
\right).
\ee
While the second term in (\ref{PSAosc}) is a single particle density,
in a future paper \cite{us} we will demonstrate that this term
describes the physical situation in which two particles are traversing
the same periodic orbit, with the same energy and are exactly half a
period out of phase. It describes the effect of particle exchange on
the spectrum and for this reason affects the symmetric and
antisymmetric spaces differently and is based on the theory of
Ref.~\cite{robbins}. Therefore, this second term also belongs to the
two-particle dynamical term and we identify (\ref{PSAosc}) as being a
purely dynamical term. We want to compare it with the corresponding
term in the symmetrized quantum densities of states. Hence, in analogy
with the previous subsection, we compare
\be\label{sclbf}
\tilde{F}^{\rm dyn}_{S/A}(L)= {\cal F} 
\left\{\tilde{\rho}^{\rm dyn}_{S/A}(k)\right\}
\ee
and
\be\label{qtmbf}
\tilde{F}^{\rm qm}_{S/A}(L)= {\cal F} \left\{
\rho_{S/A}(k)
-\bar{\rho}_{S/A}(k)
-\bar{\rho}_1*\tilde{\rho}_1(k)
\right\}.
\ee
where $\rho_{S/A}(k)$ is the quantum bosonic (S) or fermionic (A)
density of states.

\subsection{Numerics for the Symmetrized Cardioid}\label{2ppsymcardnum}

\noindent In this section, we numerically compare the symmetrized
quantum mechanics with the corresponding semiclassical quantities. In
particular, we compute the transforms (\ref{sclbf}) and (\ref{qtmbf})
\cite{comment}. The symmetrized quantum densities are
\bea \label{qtmbos}
\rho_S(k) & = & \sum_{i<j}\delta\left(k-\sqrt{k^2_i +k^2_j}\right) 
+ \sum_i \delta\left(k-\sqrt{2}k_i\right),\nonumber\\
\rho_A(k) & = & \sum_{i>j}\delta\left(k-\sqrt{k^2_i +k^2_j}\right)
\eea
using the same constraint on the energies as above. Of course, the sum
of these symmetrized densities is the total density of states.

\begin{figure}[h]
\vspace*{-3.5cm}\hspace*{-1.1cm} 
\psfig{figure=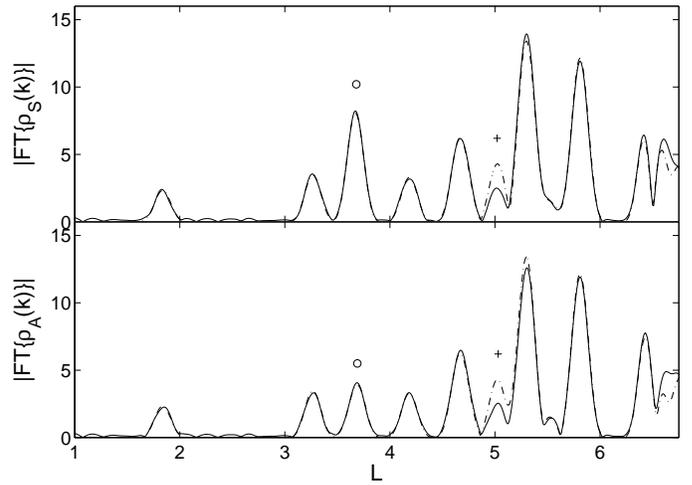,height=5.5in}
\vspace*{-3.0cm}
\caption[]{The Fourier transform of the quantum and semiclassical
symmetrized densities of states; the top is bosonic and the bottom
is fermionic. In both cases the solid line is the transform of the
quantum density of states and the dashed-dotted line is the transform
of its semiclassical approximation. Peaks with symbols are described
in the text.}
\label{cardsym}
\end{figure}

Before presenting our numerical results, we describe what we expect.
First, all the peaks of the unsymmetrised two particle density should
be present. In addition, for each periodic orbit $\Gamma$, there
should also be peaks at lengths $L_{\Gamma}/{\sqrt{2}}$ arising from
the oscillating part of the single particle density of states. The
results are shown in Fig.~\ref{cardsym}. For the two particle density
term of (\ref{sclbf}), we used the same 100 two particle orbits of
section \ref{2ppcardnum} while in the single density term we used all
single particle orbits with length $L < 11$. As well, we included the
single particle orbit ${1\over2}(^*10{\rm h})$ (not shown in
Fig.~\ref{Bruusfig}) which has a length $L=10.477$. Fig.\ref{cardsym}
displays the peak structure in the reciprocal space up to $L=6.75$.

We notice that most of the amplitude divides evenly between the
symmetric and the antisymmetric densities.  Nonetheless, there are
exceptions such as the peak at $L \approx 3.6$ (o). Here, both terms
of (\ref{PSAosc}) contribute and the difference in the sign of the
second term accounts for the uneven amplitude division between the two
symmetrized densities.  Semiclassically, we account for the peak
structure by noting that two different physical situations are
responsible for the peak structure ``o''.  First, there is the
situation in which both particles are on the orbit
$\Gamma={1\over2}(^*2{\rm a})$ with no restrictions on the time phase
difference between the two particles. This contribution comes from the
two particle density term resulting in a peak at a length
$\sqrt{2}L_{\Gamma} = 3.673$ and produces identical structures in both
densities. The second situation occurs when both particles are on the
orbit $\Gamma=(^*2{\rm a})$ exactly half a period out of phase. This
contribution comes from the single density term at
${L_{\Gamma}/\sqrt{2}}= 3.673$ and is explained more fully in
\cite{us}.  Since the second contribution comes with a different sign
in the two symmetries, the amplitudes are different for the symmetric
and antisymmetric spaces. In this particular case, it is stronger in
the symmetric density and weaker in the antisymmetric density,
although the opposite may be true in other cases.

In closing, we remark that the overall agreement between the quantum
and semiclassical calculations is good. The poorly reproduced peak
just above $L=5$ (+) comes from the single density term. This is just
the poorly reproduced peak of the single particle density
at $L \approx 7$ shifted down by a factor of $\sqrt{2}$.

\section{Conclusion}\label{conclus}

\noindent  Initially, we developed a semiclassical formalism to
describe the two particle density of states. After deriving a trace
formula describing two particle dynamics, we investigated its
structure and noted intuitive properties such as the additivity of the
actions and topological phase factors. As well, we briefly explained
the structure of the full two particle orbits which come in degenerate
families.  As a first application, we wrote down a two particle trace
formula for two identical particles in a billiard. The semiclassical
symmetry decomposition involved formal substitution of the
semiclassical quantities into the quantum mechanical expressions for
the symmetrized densities. In a future paper \cite{us}, we show how
these formal expressions emerge directly from the classical
structures.

Following these general considerations, we studied two identical
noninteracting particles in a disk and in a cardioid. In each case, we
find that the formalism correctly reproduced the full and symmetrized
densities of states. In the integrable problem, we found that our
formalism replicates the two body EBK spectrum more precisely than the
quantum spectrum, suggesting a deep connection between periodic orbit
theory and EBK quantization for integrable systems. In the chaotic
cardioid billiard, we note that the single particle orbits which pass
close to the vertex lead to inconsistencies in the Fourier transform
of the semiclassical density of states. Clearly, our formalism fails
here because the Gutzwiller theory itself fails for these
``semi-diffractive'' orbits. For all other orbits, the two particle
trace formula works very well.

The techniques employed here involve the classical phase space of each
particle. In a future paper \cite{us}, we derive the same results by
working in the full two particle phase space. This approach has the
advantage of being more general than what we have presented
here. Nonetheless, it is conceptually useful to see how the same
structure emerges from these two distinct points of view. We would
also like to incorporate interactions between the particles. Such a
project would undoubtedly require working in the full phase space
since it is no longer true that the full density of states is the
convolution of the single particle level densities. This provides an
additional motivation for working out the noninteracting problem in
the full phase space as a first step towards the more ambitious
goal. This full phase space analysis also generalises more readily to
more particles. Finally, it has the conceptual advantage that the
spurious endpoint contributions discussed in \ref{spurious_explained}
and Appendix B do not arise and therefore need not be explained away.

It may be argued that interacting many body systems are too complex to
be accessible to the semiclassical method. However, given the
intractability of the many body problem, there may well be questions
which semiclassical theory can answer. In particular, we have in mind
the applications of semiclassical theory to mesoscopic physics
\cite{houches}. Here, our seemingly academic study of billiard systems
finds physical applications in the context of nanostructures. For
example, the disk billiard can serve as a realistic lowest-order
approximation to the mean field of the electrons in a circular quantum
dot \cite{dotref}. In fact, many phenomena in ballistic mesoscopic
systems can, at least qualitatively, be described by using quantum
billiards with independent particles as physical models.

\acknowledgements{We thank Rajat Bhaduri, Matthias Brack 
and Randy Dumont for useful discussions.}

\section{Appendix A: Nonidentical Particles}

As we have mentioned, most of the discussion still applies if the two
particles are not identical. Another situation is a single particle in
a separable potential. For example, in two dimensions, one could have
$V(x,y) = V_a(x) + V_b(y)$ in which case, the dynamics in the $x$
direction are completely uncoupled from the dynamics in the $y$
direction so that the system is formally the same as if there were
distinct particles executing the $x$ and $y$ motions. The formalism
presented above follows in a natural way. The main differences are
that one no longer considers the symmetrised density of states since
the symmetry of particle exchange no longer exists and secondly there
are two distinct cross terms so that (\ref{2pdecomp}) is replaced by
\bea \label{2pdecomp_nonidentical}
\rho_2(E) &=& \bar{\rho}_{1a}*\bar{\rho}_{1b}(E)
+ \bar{\rho}_{1a}*\tilde{\rho}_{1b}(E) \nonumber\\
&+& \tilde{\rho}_{1a}*\bar{\rho}_{1b}(E) 
+ \tilde{\rho}_{1a}*\tilde{\rho}_{1b}(E),
\eea
where the indices $a$ and $b$ refer to the two distinct particles,
while the indices $1$ and $2$ still refer to one or two particle
densities of states. 

Imagine, for example, that we have two nonidentical particles in
distinct billiard enclosures. We introduce two parameters,
$\alpha_a=2m_a/\hbar^2$ and $\alpha_b=2m_b/\hbar^2$.
The smooth term (\ref{smth2pp}) is replaced by
\bea
\bar{\rho}_{1a}*\bar{\rho}_{1b}(E) &\approx& {\alpha_a \alpha_b
{\cal A}^2 \over 16\pi^2} E\nonumber\\
& - &
\left(\alpha_a^{1/2} + \alpha_b^{1/2}\right)
{\sqrt{\alpha_a\alpha_b}\over 16\pi^2}{\cal A}{\cal L}\sqrt{E}
\nonumber\\
&+& {\alpha_a^{1/2} \alpha_b^{1/2} {\cal L}^2\over {64 \pi}}
+ {(\alpha_a + \alpha_b){\cal A}{\cal K} \over 4\pi}.
\eea
The cross terms each separately have the same structure as the cross
term for identical particles. Obviously, they are no longer equal to
each other, but functionally little has changed. It is just a question
of inserting the relevant information from the different smooth and
oscillating densities of states of the two particles. Following the
same logic as before, we find
\bea \label{int_123nonid}
I_{\cal A}(E) & \approx & \phantom{-}{\alpha_a{\cal A}_a \over 4\pi^2}
\sum_{\Gamma_b} {{A_\Gamma}_b \over {T_\Gamma}_b}
\cos\left(\Phi_{\Gamma_b}
-{\pi\over 2}\right),\nonumber\\ 
I_{\cal L}(E) & \approx & -{\sqrt{\alpha_a}{\cal
L}_a\over8\pi^{3/2} \sqrt{\hbar}}\sum_{\Gamma_b} {{A_\Gamma}_b \over
\sqrt{T_\Gamma}_b} \cos\left(\Phi_{\Gamma_b} -{\pi\over
4}\right),\nonumber\\
I_{\cal K}(E) & \approx & \phantom{-}{{\cal K}_a \over \pi\hbar}
\sum_{\Gamma_b} {A_\Gamma}_b \cos\left(\Phi_{\Gamma_b}\right),
\eea
where $\Phi_{\Gamma_b}=\sqrt{\alpha_b E}L_{\Gamma_b} -
\sigma_{\Gamma_b}\pi/2$. For $\bar{\rho}_{1b}*\tilde{\rho}_{1a}(E)$,
we just interchange $a$ and $b$.

The formula for $\tilde{\rho}_{1a}*\tilde{\rho}_{1b}(E)$ still has the
same basic structure, but should obviously use the distinct periodic
orbits for particles $a$ and $b$. In particular,
Eqs.~(\ref{2pflucparts}) and (\ref{oscosc}) still apply, but with two
important differences. Firstly, the double sums over periodic orbits
are now labelled by the distinct periodic orbits of the two
particles. Secondly, the energy partition will change due to differing
masses. The criterion of stationary phase will still specify that the
two particles have the same period, but relations such as
(\ref{statcondBnonid}) and (\ref{Eo&EmEoiden}) do not apply since they
assume equal masses. The generalisations are rather straight-forward
to determine.  For example, the saddle energies (\ref{Eo&EmEoiden})
are replaced by
\bea \label{Eo&EmEononid}
{E_0\over E} = {m_a L_{\Gamma_a}^2 \over m_a L_{\Gamma_a}^2 +
m_b L_{\Gamma_b}^2},
\hspace*{0.25cm} {E - E_0\over E} = {m_b L_{\Gamma_b}^2 \over
m_a L_{\Gamma_a}^2 + m_b L_{\Gamma_b}^2}
\eea
while the general dynamical expression for billiards (\ref{ladedah})
is replaced by
\bea \label{ladedah2}
&&\tilde{\rho}_{1a}*\tilde{\rho}_{1b} (E)  \approx 
{(2E)^{3/4}\sqrt{\alpha_a\alpha_b\hbar} \over (2\pi)^{3/2}} 
\nonumber\\
&&\sum_{\Gamma_a,\Gamma_b}
{L_{\Gamma_a}L_{\Gamma_b} \over 
\left({m_aL_{\Gamma_a}^2+m_bL_{\Gamma_b}^2}\right)^{5/4}}
A_{\Gamma_a}(E_0)A_{\Gamma_b}(E-E_0)\nonumber\\
&&\cos \left(
\sqrt{\alpha_aL_{\Gamma_a}^2+\alpha_bL_{\Gamma_b}^2}\sqrt{E} - 
\left(\sigma_{\Gamma_a} + \sigma_{\Gamma_b}\right) {\pi \over 2} -
{\pi\over 4} \right).
\eea
In the special case of identical particles, it is simple to check
that this expression reduces to (\ref{ladedah}).
For lack of an immediate physical context, we do not explore this case
any further.

\section{Appendix B: Spurious End-point Contributions for the Cardioid}

Here we evaluate the cross term integrals exactly for isolated
periodic orbits. This allows us to do an asymptotic expansion to
explicitly demonstrate that the additional endpoint contributions not
included are spurious. We must evaluate the integral
\be\label{convcrossbill}
\bar{\rho}_1 * \tilde{\rho}_1 (E) = \int_{0}^{E}\d\eps
\bar{\rho}_1(\eps) \tilde{\rho}_1 (E - \eps)
\ee
where $\bar{\rho}_1(\eps)$ is given by the Weyl expansion
(\ref{pave1pbill}) and $\tilde{\rho}_1(\eps)$ for a billiard with
isolated orbits is given by
\be\label{osc1pbill}
\tilde{\rho}_1(\eps) \approx {\alpha^{1/2} \over {2 \pi
\sqrt{\eps}}}  \sum_{\Gamma} {{{L_{\gamma}}} \over
{\sqrt{\left| \det \left( \tilde{M}_{\Gamma} - I \right) \right|}}}
\cos \left(\sqrt{\alpha\eps} L_{\Gamma} -
\sigma_{\Gamma}{\pi \over 2}\right).
\ee
This gives
\bea\label{crossbill}
\bar{\rho}_1*\tilde{\rho}_1 (E) & \approx &\sum_{\Gamma}
{{{L_{\gamma}}} \over {\sqrt{\left| \det \left( \tilde{M}_{\Gamma} - I
\right) \right|}}} \nonumber\\
&&\left(\alpha^{3/2} {{\cal A} \over {8
\pi^2}} I_1 - \alpha {{\cal L} \over {16
\pi^2}} I_2 + \alpha^{1/2} {{{\cal K}} \over {2 \pi}} I_3
\right)
\eea
where
\bea\label{crossint1}
I_1 & = & \int_{0}^{E} \d\eps {1 \over \sqrt{E-\eps}} 
\cos\left(\sqrt{\alpha(E-\eps)} L_{\Gamma} -
\sigma_{\Gamma}{\pi \over 2}\right),\nonumber \\
I_2 & = & \int_{0}^{E}\d\eps {1 \over \sqrt{\eps}}{1 \over
\sqrt{E-\eps}} \cos \left(\sqrt{\alpha(E-\eps)}
L_{\Gamma} - \sigma_{\Gamma}{\pi \over 2}\right),
\nonumber \\
I_3 & = & {1 \over \sqrt{E}} \cos \left(\sqrt{\alpha E}
L_{\Gamma} - \sigma_{\Gamma}{\pi \over 2}\right).
\eea
If we evaluate the first two integrals exactly, we get 
\be\label{crossint1exact}
I_1={2\over{\alpha^{1/2} L_\Gamma}} \left(
\cos\left(\Phi_\Gamma + \phi_\Gamma -{\pi\over2}\right)
-\cos\left(\phi_\Gamma-{\pi\over2}\right)\right)
\ee
and
\bea\label{crossint2exact}
I_2 & = & \pi\cos\phi_\Gamma \; J_0\left(\Phi_\Gamma\right) 
- \pi\sin\phi_\Gamma \;{\bf H}_0\left(\Phi_\Gamma\right)\nonumber\\
& \approx & \sqrt{2\pi\over\sqrt{\alpha E}L_\Gamma}
\cos\left(\Phi_\Gamma + \phi_\Gamma - {\pi\over4}\right)\nonumber\\
&& - {2\over\sqrt{\alpha E}L_\Gamma}\cos\left(\phi_\Gamma\right) + \cdots
\eea
where $\Phi_\Gamma=\sqrt{\alpha E} L_\Gamma$,
$\phi_\Gamma=-\sigma_\Gamma\pi/2$, $J_{0}$ is a zero-order Bessel
function and ${\bf{H}}_{0}$ is a zero-order Struve function. In the
second line of (\ref{crossint2exact}), we have used the asymptotic
expansions of these two functions.

In both $I_1$ and $I_2$, we note that asymptotically there are terms
with two distinct structures. The first are terms which are sinusoidal
in $\sqrt{E}$ and correspond exactly to what was used as the cross
term for the cardioid ({\it i.e.} Eqs.~(\ref{int_1a}) and
\ref{int_23})). There are also terms which are nonsinusoidal in 
$E$. In $I_1$, this comes directly from the upper endpoint of the
integral while in $I_2$ it comes from the expansion of the Struve
function.  In each term, the nonsinusoidal terms arise from the
endpoint around $\eps=E$ which, as we argued in section
\ref{spurious_explained}, corresponds to an unphysical
situation. Therefore, keeping only the asymptotically appropriate term
({\it i.e.} the oscillatory one) yields the correct behaviour for the
cross term.

A similar analysis would yield similar results for the spurious
endpoint contributions in the cross term of the disk billiard and the
dynamical term of either billiard.


\begin{references}
\bibitem{comment} We use different window function parameters for
the single and two particle densities in Eqs.~(\ref{sclbf}) and
(\ref{qtmbf}), although this is not strictly necessary. For single
particle densities, the parameters $k_0$ and $k_f$ are chosen so the
window function goes smoothly to zero at the endpoints of the single
particle spectrum. For two particle densities, we use the same window
function as before.
\end{references}
\end{document}